\documentclass[manuscript=review,utf8x]{beilstein}

\usepackage{xspace}

\usepackage{amsmath,amssymb}
\usepackage[]{color}
\usepackage[]{float}
\usepackage[version=3]{mhchem}
\usepackage[]{graphicx}
\usepackage[]{hyperref}

\newcommand{\dd}{{\rm d}}

\newcommand{\intqmqp}{\int_{q_-}^{q_+}}

\newcommand{\qb}{\textbf{q}}

\newcommand{\qw}{(\textbf{q},\omega)}

\newcommand{\rb}{\textbf{r}}

\newcommand{\rt}{(\textbf{r},t)}

\newcommand{\vb}{\textbf{v}}

\newcommand{\vu}{\hat{\textbf{v}}}

\begin{document}


\title{Surface excitations in the modelling of electron
transport for electron-beam-induced deposition experiments}
\author*{Francesc Salvat-Pujol}{salvat-pujol@itp.uni-frankfurt.de}
\affiliation{Institut f\"ur Theoretische Physik, Goethe-Universit\"at Frankfurt, Max-von-Laue-Stra{\ss}e 1, 60438 Frankfurt am Main, Germany}
\author[1]{Roser Valent\'{i}}
\author{Wolfgang S.\ Werner}
\affiliation{Institut f\"ur Angewandte Physik, Technische Universit\"at
Wien, Wiedner Hauptstra{\ss}e 8-10/134, 1040 Wien, Austria}
\maketitle
\begin{abstract}
  The aim of the present overview article is to raise awareness of an
  essential aspect that is usually not accounted for in the modelling of
  electron transport for focused-electron-beam-induced deposition (FEBID) of
  nanostructures: surface excitations are on the one hand responsible
  for a sizeable fraction of the intensity in
  reflection-electron-energy-loss spectra for primary electron energies
  of up to a few keV and, on the other hand, they play a key role in the
  emission of secondary electrons from solids, regardless of the primary
  energy. In this overview work we present a general perspective of
  recent works on the subject of surface excitations and on low-energy
  electron transport, highlighting the most relevant aspects for the
  modelling of electron transport in FEBID simulations.
\end{abstract}


\section{Introduction}

An accurate modelling of the energy losses of electrons traversing a
solid surface is instrumental for a quantitative understanding of a
series of techniques exploiting transmitted, reflected, or emitted
electrons, including a number of spectroscopies (electron-energy-loss
spectroscopy, x-ray photoelectron spectroscopy, and Auger-electron
spectroscopy), electron microscopy, and the
focused-electron-beam-induced deposition (FEBID) of nanostructures, on
which we focus here. This technique employs beams of focussed
keV-electrons to trigger and steer the growth of nanostructures with
tunable electronic and magnetic properties from molecules of
organometallic precursor gases \cite{huth2012} adsorbed on a substrate
\cite{utke2008}. It has been shown that, for irradiation with electrons
of 1-5 keV, both the incoming primary electrons and the emitted
secondary electrons influence the growth of the nanostructure, the
latter electrons being responsible for the lateral resolution
\cite{smith2007}.

In the modelling of electron transport for FEBID
\cite{silvis-cividjian2005, liu2006, smith2007, smith2008, utke2008,
salvat-pujol2013ebid}, electron stopping is described on the basis of
properties that are applicable in the bulk of the material.  However,
electrons traversing a solid interface additionally excite surface modes,
an energy-loss channel that amounts to a sizeable fraction of the
energy-loss spectrum for electrons of up to a few keV.  The existence of
surface excitations was predicted in the late 1950s \cite{ritchie1957};
experimental evidence was obtained shortly thereafter
\cite{powell1959al,powell1959mg}. In order to model surface excitations
in electron spectroscopies, several models have been developed to date
\cite{ritchie1966, geiger1967, otto1967, kroeger1968, kroeger1970,
chan1973, jacob1973, chan1975i, chan1975ii, chan1976, arista1994,
tung1994, chen1996dimfpsurf, yubero1996, denton1998, ding1998i,
ding1998ii, kwei1998, vicanek1999, li2005, salvat-pujol2013}.  Various
approaches are considered, often with underlying simplifications, evoked
on physical or technical grounds, in the interest of making calculations
feasible in a finite time. In order to derive a distribution of energy
losses of charged projectiles moving in the vicinity of the surface,
some of the models cited above rely on the semiclassical dielectric
formalism, whereas others adopt a many-body formalism. Both approaches
have been shown to yield results in equivalently good agreement
\cite{da2011} with experimental data.

In what follows we briefly review the stopping of charged projectiles in
the vicinity of a solid surface, along the lines of
\cite{salvat-pujol2013}, which will be referenced for further details.
We summarize a series of rules which characterize the behavior of the
probability for surface excitations and we briefly review a practical
model for the emission of secondary electrons. Relevant aspects to FEBID
modelling will be highlighted.

\section{Inelastic collisions in the bulk of the material}

Energy losses of a charged projectile moving in a solid can be described
accurately within the semiclassical dielectric formalism. In this
approach, one assumes that the presence of the charged projectile
disturbs the equilibrium charge density of the solid, which becomes
polarized and, thus, an electric field is induced at all points of
space. The force acting on the charged projectile due to the induced
electric field is assumed to be the agent responsible for its
(electronic) stopping. In order to derive physical quantities that
describe the stopping, it is now a matter of calculating first the
induced electric field and, from it, the so-called stopping power,
defined as the variation of the kinetic energy of the projectile per
unit path length. Once an expression for the stopping power is derived,
one can identify from it an expression for the distribution of energy
losses per unit path length, the basic quantity that is needed to
describe energy losses in a detailed Monte-Carlo simulation of electron
transport. In this section we briefly outline the basic steps of these
calculations and highlight the underlying assumptions. Further details
can be found in the cited works.

The starting point of the calculation is the dielectric function
$\varepsilon(q,\omega)$ of the material, where $q$ and $\omega$ are the
respective Fourier-conjugate variables of the position, $\rb$, and the
time, $t$. In practice one typically has data available for
$\varepsilon(\omega)$, be it from optical data obtained experimentally
\cite{werner2009opt} or from theoretical calculations, \textit{e.\ g.},
via density-functional theory calculations
\cite{koepernik1999,ambrosch-draxl2006,werner2009opt}. An
$\omega$-dependent dielectric function is sufficient to describe the
response of the medium to a spatially homogeneous perturbation, such as
that of an incoming photon. However, for incoming charged projectiles
the perturbation is strongly dependent on the spatial coordinates, so
that a $q$-dependent dielectric functions is required. Physically
reliable models are built on the basis of the $(q,\omega)$-dependent
dielectric function for the homogeneous electron
gas \cite{lindhard1954,mermin1970,denton2008} or on the basis of a simple
superposition of Drude-Lindhard oscillators \cite{werner2009opt}.

Assuming a projectile that moves with a velocity $\vb$ along a
trajectory $\rb=\vb t$, one can conveniently solve the Maxwell equations
in Fourier space to obtain the following expression for the induced
electric field \cite{salvat-pujol2013}
\begin{equation}
  \textbf{E}_\textrm{ind}\qw
  =
  -\textrm{i}4\pi
  \frac{\mathbf{q}}{q^2}
  \rho(\mathbf{q},\omega)
  \left[
    \frac{1}{\varepsilon(\mathbf{q},\omega)}
    -1
  \right],
  \label{eq:eind}
\end{equation}
where $\rho\qw$ is the Fourier transform of the projectile charge
density $\rho\rt=Z_0e\delta(\rb-\vb t)$, where $Z_0$ is the projectile
charge in units of the modulus of the electron charge, $e$, and $\vb$ is
the velocity of the projectile. To obtain this expression, the following
approximations were considered: (1) the Coulomb gauge was adopted and
the contribution from the vector potential was neglected (2) the
dielectric displacement field was assumed to be proportional to the
electric field in Fourier space (linear response). The first
approximation restricts the validity of the calculation to
non-relativistic projectiles (the calculation with the full electric
field for relativistic projectiles is also feasible
\cite{schattschneider2005}), whereas the second approximation can be
seen to be formally equivalent to a first-order Born approximation in
perturbation theory, imposing a lower bound to the domain of validity of
the calculation \cite{tung1994,bote2008}, which for practical purposes
is above 100 eV.

The stopping power $S$ is obtained as the variation of the kinetic
energy per unit path length,
\begin{equation}
  S
  =
  -
  \frac{\dd \mathcal{E}}{\dd s}
  =
  \left.
    \vu
    \cdot
    \mathbf{E}_\textrm{ind}(\rb,t)
  \right|_{\rb=\vb t}
  ,
  \label{eq:S}
\end{equation}
where $\mathcal{E}$ is the kinetic energy of the projectile and $s=vt$
is the path length. Combining Eqs.\ (1) and (2) one obtains
\begin{equation}
  S
  =
  \frac{2(Z_0)^2}{\pi}
  \frac{1}{v^2}
  \int_0^\infty \textrm{d}q\;
    \frac{1}{q}
  \int_{0}^{qv} \textrm{d}\omega\;
    \omega\;
    \textrm{Im}
    \left[
      -\frac{1}{\varepsilon(q,\omega)}
    \right]
  .
  \label{}
\end{equation}
Up to this point the stopping of the projectile is treated as a
continuous phenomenon, whereas in reality charged projectiles lose
energy and are deflected in the course of individual inelastic
collisions. The so-called semiclassical approximation consists in
assigning to $\hbar \qb$ and $\hbar \omega$ the meaning of a momentum
transfer from the projectile to the medium and of an energy loss of the
projectile, respectively. Atomic units ($\hbar=m_e=e=1$) will be used
below. Now that these variables have a well-defined physical meaning,
the corresponding integrals must be restricted to the kinematically
allowed domain,
\begin{equation}
  S(\mathcal{E})
  =
  \frac{2(Z_0)^2}{\pi}
  \frac{1}{v^2}
  \int_0^\mathcal{E} \textrm{d}\omega\;
    \omega\;
  \intqmqp \textrm{d}q\;
    \frac{1}{q}
    \textrm{Im}
    \left[
      -\frac{1}{\varepsilon(q,\omega)}
    \right]
    ,
  \label{eq:s0}
\end{equation}
where
\begin{equation}
  q_\pm
  =
  \sqrt{2\mathcal{E}}
  \pm
  \sqrt{2(\mathcal{E}-\omega)}
  \label{}
\end{equation}
are the minimum (-) and maximum (+) allowed momentum transfers allowed
by the energy and momentum conservation laws.

Equation (4) can be understood as the average energy loss per unit path
length dictated by a distribution of energy losses per unit path length,
$\dd\mu/\dd \omega$:
\begin{equation}
  S(\mathcal{E})=
  \int_0^\mathcal{E} \textrm{d}\omega\;
    \omega\;
    \frac{\dd\mu}{\dd\omega}
  .
  \label{}
\end{equation}
The quantity $\dd\mu/\dd\omega$ is known as the differential inelastic
inverse mean free path (DIIMFP), explicitly given by
\begin{equation}
  \frac{\dd\mu}{\dd\omega}
  =
  \frac{2(Z_0)^2}{\pi}
  \frac{1}{v^2}
  \intqmqp \textrm{d}q\;
    \frac{1}{q}
    \textrm{Im}
    \left[
      -\frac{1}{\varepsilon(q,\omega)}
    \right]
  .
  \label{}
\end{equation}
Note that the DIIMFP is a function of the energy loss for the given
velocity of the projectile. The integral of the DIIMFP over all allowed
energy losses gives the inelastic inverse mean free path
\begin{equation}
  \lambda_i^{-1}
  =
  \int_0^\mathcal{E}\dd\omega\;
    \frac{\dd\mu}{\dd\omega}
  .
  \label{}
\end{equation}
The latter two quantities are the necessary quantities for a detailed
Monte Carlo simulation of electron transport (see section ``Monte-Carlo
simulation of electron energy-loss spectra''), a method that has been
successfully used in the last decades. 

\section{Inelastic collisions in the vicinity of a planar surface}

The scheme outlined in the previous section to describe inelastic
interactions of charged projectiles in solids gives a good account of
inelastic collisions in the bulk of the solid. However, projectiles
impinging and emerging from a solid additionally cross a planar
interface to vacuum (or another solid) that is not explicitly accounted
for. The existence of a plane surface imposes additional boundary
conditions on the electric field \cite{jackson,griffiths}. 

Several approaches exist in the literature to solve the Maxwell
equations with these boundary conditions for the stopping problem: some
consider the dielectric function of a semi-infinite medium
\cite{nazarov1995}, and others (preferred in the electron-spectroscopy
community) rely on a method which allows one to work with bulk
dielectric functions, the method of image charges, also known as the
method of extended pseudomedia. The method consists in rephrasing the
semi-infinite-geometry problem as the sum of two infinite-geometry
problems, supplied with a series of fictitious charges that are
determined in terms of known quantities by imposing the boundary
conditions at the interface. 

The resulting induced electric field has a more complex expression than
in the bulk case. Nevertheless, it can be expressed as the sum of one
contribution arising from a charge density induced in the bulk of the
material and another one arising from a charge density induced at the
surface of the material.

The DIIMFP resulting from the induced electric field becomes more
complicated, with two additional parametric dependencies: (1) on
the depth coordinate with respect to the surface and (2) on the surface
crossing angle with respect to the surface normal. Several models exist
with varying approximations
\cite{yubero1996,chen1996dimfpsurf,li2005,salvat-pujol2013}, the effect
of which was scrutinized \cite{salvat-pujol2013}. Regardless of the
details of the models, they all yield a number of consistent general
features and trends of the surface excitation probability:
\begin{itemize}
  \item Surface energy losses can be undergone by the charged projectile
    on either side of the interface, at the solid side or at the vacuum
    side (!). Indeed, a surface charge can be induced regardless of the
    side at which the projectile is moving on and, thus, a charged
    projectile moving on the vacuum side of the interface can also
    undergo energy losses. It has been recently shown that, in
    reflection-electron-energy-loss spectra, surface losses on the
    vacuum side of the interface account for a large fraction of the
    surface-excitation intensity, often more than half of
    it \cite{werner2013ss}.

  \item The probability for an electron that crosses a surface to
    undergo a surface excitation is, to a first
    approximation \cite{werner2001threestep}, proportional to the surface
    dwell time $t\sim1/(\sqrt{\mathcal{E}}\cos\theta)$, where
    $\mathcal{E}$ is the projectile energy and $\theta$ is the surface
    crossing angle with respect to the surface normal. The energy
    dependency implies that, in practice, surface excitations are
    relevant for electron energies up to a few keV. Additional structure
    to the aforementioned angular behavior is predicted for scattering
    geometries coinciding with deep minima of the differential elastic
    scattering cross section: minor deflections in the course of an
    inelastic collision lead to an effective scattering geometry with
    enhanced elastic scattering and therefore higher detection
    probability \cite{werner2013ss}.

  \item The DIIMFP for energy losses of charged projectiles impinging on
    a surface differs from the DIIMFP for the conjugate emerging
    direction. This effect, known as in-out asymmetry in surface
    energy-losses, has been long predicted but only recently observed
    experimentally \cite{salvat-pujol2014inout}. In-out differences are
    most accentuated for surface-crossing directions close to the
    surface normal and for high kinetic energies ($\sim1$ keV).

\end{itemize}

\section{Monte-Carlo simulation of electron energy-loss spectra}
\label{sec:mc}

\begin{figure}
  \centering
  \includegraphics[scale=1]{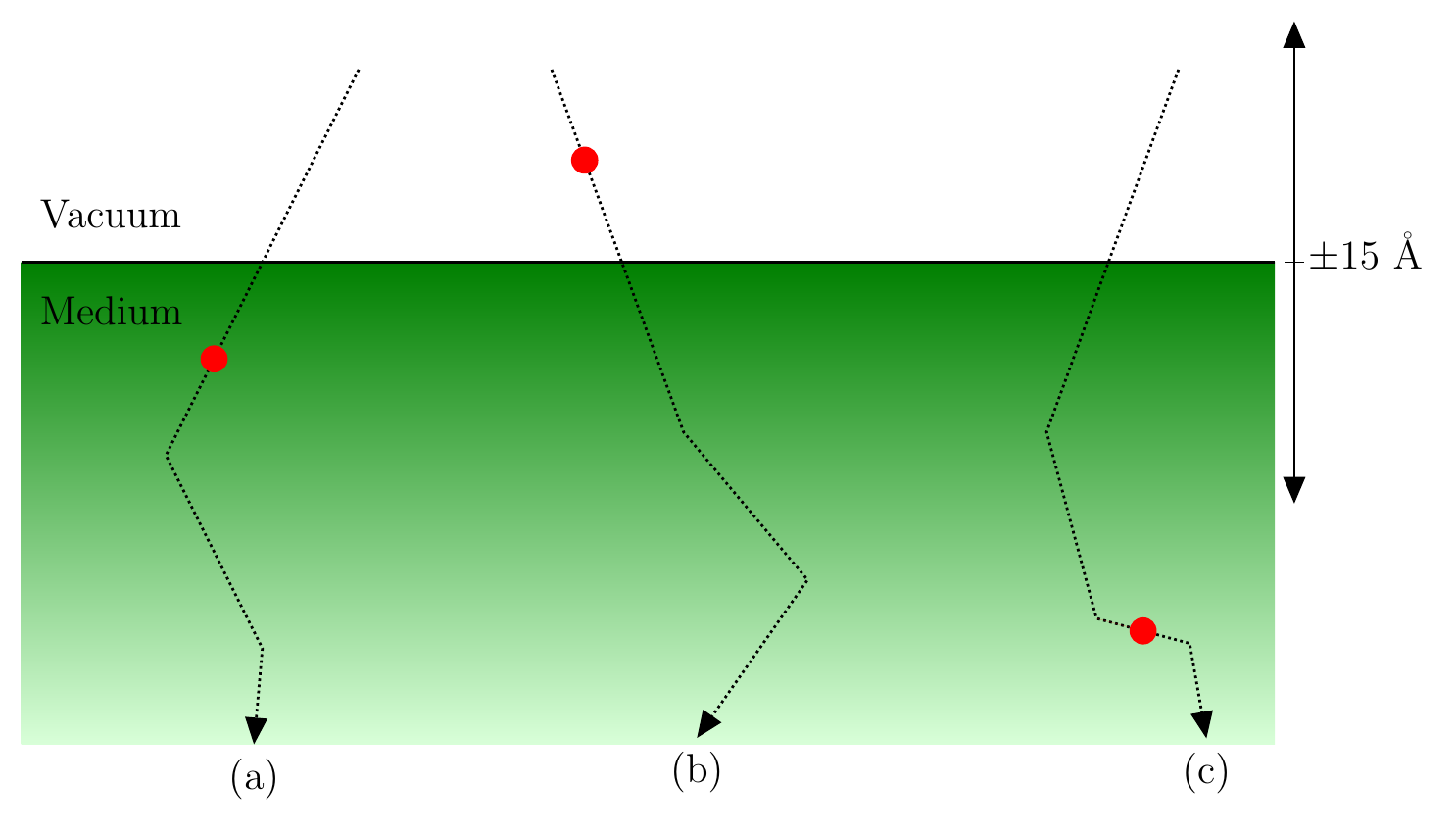}
  \caption{Example incoming trajectories (dotted lines) in the
    surface-scattering zone (typically $\pm15$~\AA\ about the surface),
    undergoing (a) a surface energy loss in the medium side, (b) a
    surface energy loss in the vacuum side, (c) a bulk energy loss,
    indicated by the filled circles.}
  \label{fig:surf}
\end{figure}

The electron-transport problem in a solid is described in terms of a
Boltzmann-type transport equation. A practical method for solving the
problem is provided by Monte-Carlo simulation, which consists in
sampling an ensemble of trajectories undergoing collisions of the
relevant types as dictated by a given set of interaction cross sections.
A statistical average of the desired observable is performed over the
sampled trajectories to the selected precision
\cite{werner2001tutorial}.

\begin{figure}
  \centering
  \includegraphics[scale=1]{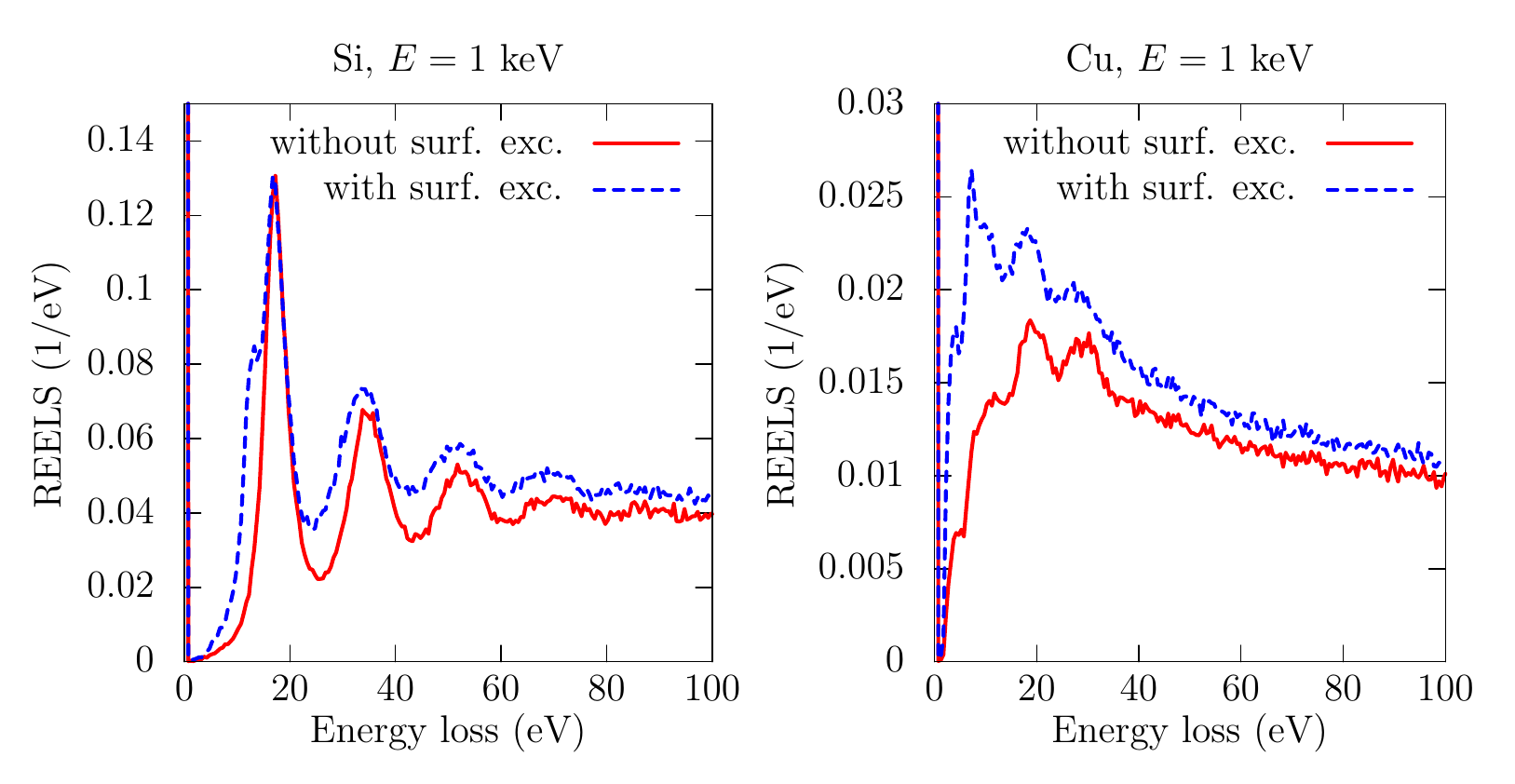}
  \caption{Comparison of reflection-electron-energy-loss spectra (REELS)
    of Si (left) and Cu (right) under bombardment with 1-keV electrons
    at normal incidence, without (red solid curves) and with (blue
    dashed curves) an account of surface excitations. All backscattered
  electrons are collected.} \label{fig:reels1000}
\end{figure}

\begin{figure}
  \centering
  \includegraphics[]{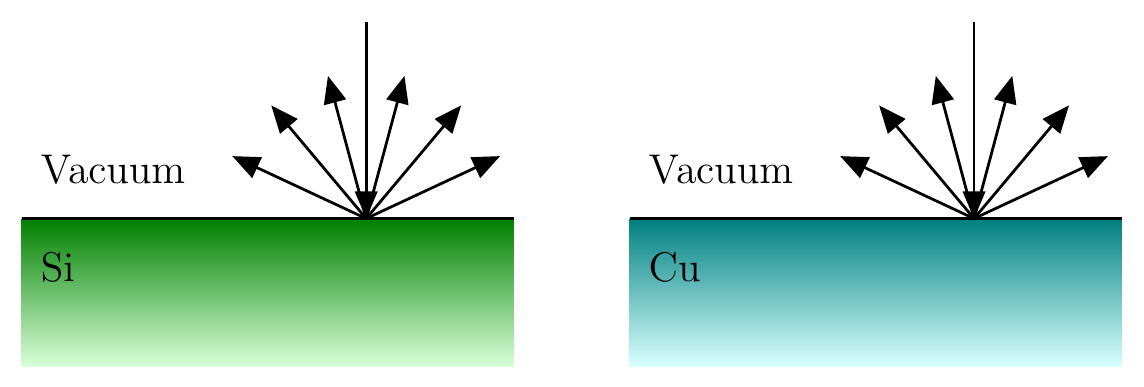}
  \caption{Simulation geometry: 1-keV electrons impinge normally onto
  the material (Si or Cu); all backscattered electrons are collected.}
  \label{fig:geom}
\end{figure}

In the energy range between 100 eV and a few keV the relevant
interaction mechanisms of electrons with the solid are elastic
collisions with the atoms and inelastic collisions with typically weakly
bound electrons in the solid. Elastic scattering can be accurately
described by means of a differential cross section for elastic
scattering (DCES), which can be systematically calculated by means of
partial-wave calculations \cite{salvat2005,bote2009}. Inelastic
scattering is accounted for by the DIIMFPs described above. Monte-Carlo
simulations of electron transport (bulk losses only) for typical
geometries in FEBID experiments have been previously considered
\cite{utke2008,salvat-pujol2013ebid}. The inclusion of surface
excitations implies a modification of the sampling algorithm in the
vicinity of the surface (typically $\pm 15$\AA\ about the surface), as
schematically shown in Fig.\ \ref{fig:surf}.
Technical details on the implementation of the algorithm for the
simulation of surface energy losses can be found elsewhere in great
detail \cite{novak2008,li2005,salvat-pujol2013}. Here the focus is on
the effect of surface excitations on the reflection-electron-energy-loss
spectrum (REELS). To this effect, Fig.\ \ref{fig:reels1000} compares the
REELS of Si (left) and Cu (right) under bombardment with 1-keV electrons
impinging perpendicularly on the sample; all reflected electrons are
collected. The simulation geometry is depicted in Fig.\ \ref{fig:geom}.
The materials are chosen as representative substrate (Si) and deposit
(Cu) materials. The solid red curves (dashed blue curves) in Fig.\
\ref{fig:reels1000} correspond to REELS simulated without (with) the
inclusion of surface excitations. We observe that even for a primary
energy of 1 keV surface excitations account for (1) additional features,
\textit{i.\ e.} the excitation of surface plasmons, in the
low-energy-loss part of the REELS that are not accounted for by a
bulk-only description of the energy losses of charged projectiles in the
material and (2) a sizeable fraction of the intensity in the first few
tens of eV of energy losses, about 20\% of the intensity in the case of
Si and 15\% of the intensity in the case of Cu.  Although the relative
importance of surface excitations is enhanced for lower energies, their
effect is noticeable even in the 1-keV domain.  Thus, the inclusion of
surface excitations in the modelling of electron-transport is expected
to give a yet more quantitative description of FEBID processes at and
below the 1-keV primary-energy domain.

\section{Secondary-electron emission}

\begin{figure}
  \centering
  \includegraphics[width=0.5\textwidth]{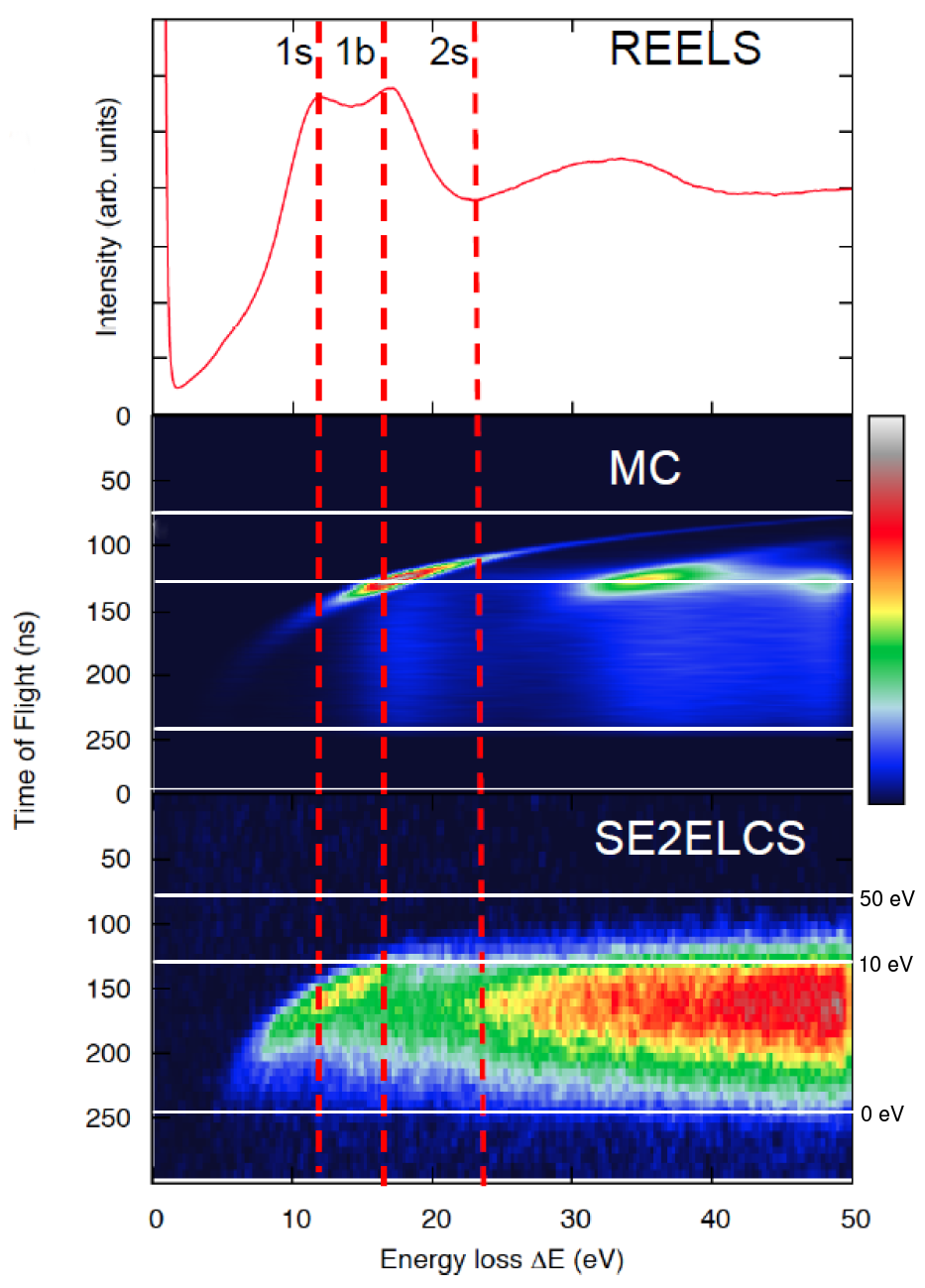}
  \caption{(Upper panel) Reflection-electron-energy-loss spectrum
    (REELS) of Si under 100 eV bombardment (see \cite{werner2013} for
    experimental details). (Lower panel) (e,2e)-coincidence spectrum of
    secondary electrons emitted in coincidence with energy losses
  (SE2ELCS) of 100-eV electrons backscattered from Si. (Middle panel)
Monte-Carlo simulation of the SE2ELCS measurement without accounting for
surface energy losses. The vertical dashed lines in red indicate energy
losses corresponding to the excitation of one surface, one bulk, and two
surface plasmons. The horizontal solid white lines indicate the times of
flight corresponding to electrons with 0 eV (accelerating grids were
used), 10 eV, and 50 eV.}
  \label{fig:e2esinosurf}
\end{figure}

Energy losses of the charged projectile can lead to the ejection of
loosely bound electrons of the solid, which emerge as secondary
electrons (SE). The majority of these SE are of relatively low energies
($\leq 50$~eV). These energies are well below the domain of validity of
the elastic and inelastic interaction cross sections available in the
literature, which has been a limitation for progress in the field.
Electron coincidence measurements
\cite{voreades1976,pijper1991,muellejans1992,muellejans1993} have
supplied a wealth of valuable information. Recently, coincidence
measurements of correlated electron
pairs emitted from solids (Al, Si, Ag) under electron bombardment have
been measured, providing a double-differential SE yield, differential
with respect to the energy loss of the primary electron and with respect
to the energy (or the time of flight) of the emitted secondary electron
\cite{werner2013}. These experimental data are displayed for Si under
100-eV electron bombardment in the lower panel of Fig.\
\ref{fig:e2esinosurf} as a
bird's-eye-view plot (only the shape and relative intensities of the
spectrum are of relevance here, hence the missing units in the linear
color scale, where black is the null point and white is the maximum
attained value). The horizontal white lines indicate the corresponding
times of flight for electrons with 0 eV (accelerating grids were used),
50 eV, and 100 eV. See \cite{werner2013} for the experimental details.
The plot can be read as the (time-of-flight) spectrum of secondary
electrons emitted as a result of different energy losses of the
impinging electron (to be read at the abscissae). The upper panel of
Fig.\ \ref{fig:e2esinosurf} displays the REELS of 100 eV from Si, where the energy-loss
peaks corresponding to the excitation of one surface plasmon, one bulk
plasmon, and two surface plasmons are indicated by vertical red dashed
lines and labeled, respectively, 1s, 1b, 2s as a guideline for the
abscissae scale in the other plots of the figure.

\begin{figure}
  \centering
  \includegraphics[width=0.5\textwidth]{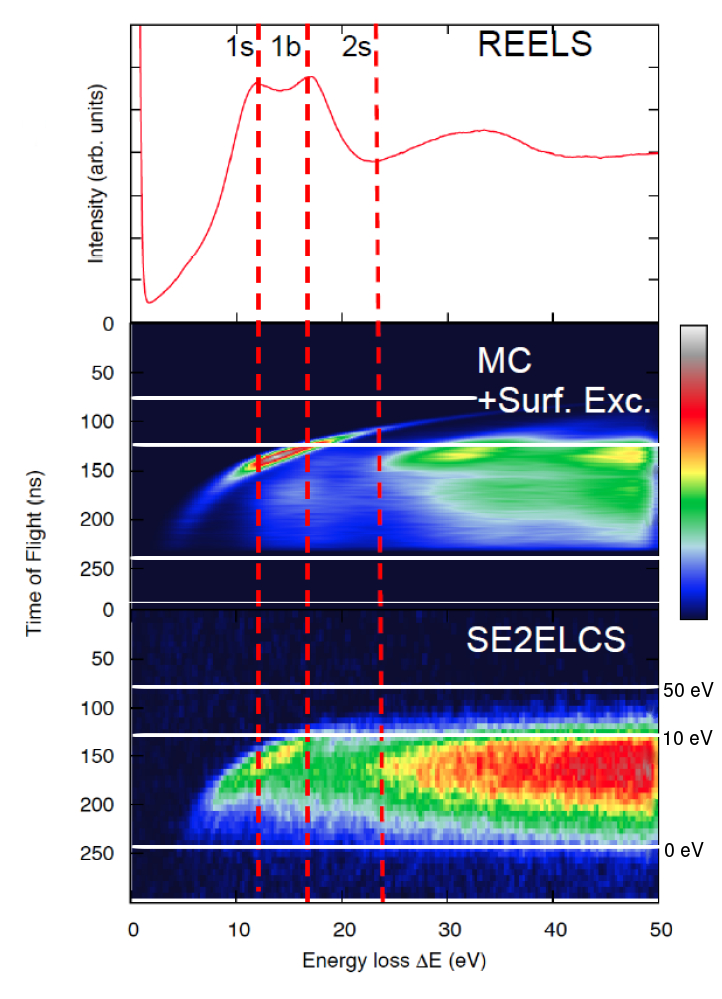}
  \caption{Same as Fig.\ \ref{fig:e2esinosurf} with the inclusion of surface excitations in
  the modelling of electron transport through the solid-vacuum interface.}
  \label{fig:e2esisurf}
\end{figure}

The coincidence data (\textit{e.\ g.}, lower panel of Fig.\
\ref{fig:e2esinosurf}) provide
on the one hand very detailed insight into the mechanisms responsible
for SE emission and, on the other hand, provide a benchmark
against which models for SE emission and low-energy electron transport
in general can be tested. The transport models, in turn, aid in the
interpretation of the data, as discussed below. The Monte-Carlo
simulation briefly outlined above was extended to include the generation
and the transport of the secondary electrons and to simulate the
electron-coincidence measurement on the basis of a simple model for SE
emission: every time that the primary electron undergoes an energy loss,
a SE trajectory is started with the energy loss as an initial energy
(see \cite{werner2013} for the simulation details). Having the
experimental data as a guideline, the interaction cross sections
described above were used down to 1 eV (knowing that this is well below
the domain where they are formally applicable) as a first approximation.
Simulations were first carried out using bulk energy-loss DIIMFPs
exclusively. The resulting spectrum is displayed in the middle panel of
Fig.\ \ref{fig:e2esinosurf}. It is clear that these simulated peaks do not
reproduce the onset of the experimentally observed peaks. Only after the
inclusion of surface excitations, both for the incoming primary
electrons, for the backscattered electrons, and for the emitted
secondary electrons, is good agreement between simulations and
measurements found, as shown in Fig.\ \ref{fig:e2esisurf}. The Monte Carlo simulations further
allow one to discern the processes that give rise to the
different regions of the coincidence spectrum \cite{werner2013}.

Thus, it was found that any realistic model of SE emission and
low-energy electron transport near solid surfaces must  account for
surface excitations. This conclusion has strong implications on the emission depth
from which SE are emitted: if secondary electrons undergo additional energy losses on
their way out of the solid, the average SE-emission depth becomes much shallower than
one would assume on the basis of a model based only on bulk properties.
The predicted number of emitted SE can also differ
appreciably with respect to a bulk-only simulation. Furthermore, the
energies of the SE are also modified by the presence of additional
surface energy-loss channels.

\section{Conclusions}

In light of the presented richness in the behavior of surface
excitations and their effect on both the energy losses of the impinging
electrons and on the emission of secondary electrons, it is to be
expected that their inclusion in the modelling of electron transport for
FEBID will yield a more detailed description of the role played by both
the primary electrons and the emitted secondary electrons in the growth
process. It should be noted that, while surface excitations are relevant
for primary electrons with energies up to 1-2 keV, they are essential
ingredients for the modelling of slow secondary electrons regardless of
the energy of the primary electron responsible for their emission.

The previous considerations suggest that the inclusion of surface
excitations in the electron-transport model employed to investigate
FEBID experiments might lead to noticeable effects. On the one hand,
more primary electrons are backscattered compared to the case without
surface excitations (see Fig. 2), so that an increase in the simulated
deposition rate might be expected (at least for primary energies in the
1-2 keV regime and below). On the other hand, more slow (<=~50 eV)
secondary electrons will be available from the decay of surface plasmons
\cite{chung1977} excited by either the incoming electrons or the outgoing
electrons (backscattered electrons or emitted secondary electrons). This
should also contribute to an increase of the simulated deposition rate
and additionally lead to an enhancement of the FEBID proximity effect.

\acknowledgements

F.\ S.-P.\ acknowledges the support of the Alexander von Humboldt
Foundation through a Humboldt Research Fellowship. This work was
conducted within the framework of the COST Action CM1301 (CELINA).
Financial support by the Austrian Science Fund FWF (Project No.
P20891-N20) is gratefully acknowledged.

\bibliography{refs}

\end{document}